\documentclass[twoside,11pt]{article}
\usepackage{a4wide,cite,latexsym,amsfonts,amssymb,exscale,epsfig}
\usepackage[centertags,sumlimits,intlimits,namelimits,reqno]{amsmath}

\def\theauthor{}
\def\empty{}
\def\theaffiliation{}
\def\preprint#1{
  \thispagestyle{plain}
  \def\theauthor{#1}
  \ifx\theauthor\empty
  \else
    \begin{flushright}{\small #1\par}\end{flushright}
  \fi
  \begin{center}}
\def\title#1{
  {\LARGE #1\par}\vskip 1em}
\def\author#1{
  \ifx\theaffiliation\empty
  \else
    \par
  \fi
  \def\theauthor{#1}\def\theaffiliation{}}
\def\email#1{
  \vskip 1em{\large\theauthor\footnote{\small email: {\tt #1}}\par}\vskip
.5em}
\def\affiliation#1{
  \ifx\theaffiliation\empty
    \def\theaffiliation{second}
  \else
  \fi
  {\small\sl #1\par}}
\def\date#1{
  \vskip 1em{(#1)\par}\end{center}\vskip 2em}
\def\maketitle{
}

\def\acknowledgments{\section*{Acknowledgments}}%
\def\pacs#1{\noindent {\small PACS: #1\par}}%
\def\keywords#1{\noindent {\small keywords: #1\par}}%

\usepackage{bbm}

\def\R{{\mathbbm R}}
\def\Z{{\mathbbm Z}}

\def\SO{{SO}}

\def\SU{{SU}}

\def\U{{U}}

\def\g{{\mathfrak{g}}}
\def\h{{\mathfrak{h}}}

\def\so{{\mathfrak{so}}}

\def\su{{\mathfrak{su}}}

\def\ie{{\sl i.e.\/}}

\def\etc{{\sl etc.\/}}
\def\cf{{\sl c.f.\/}}
\let\phi=\varphi
\let\theta=\vartheta
\let\epsilon=\varepsilon

\def\tr{\mathop{\rm tr}\nolimits}

\def\Aut{\mathop{\rm Aut}\nolimits}

\def\Der{\mathop{\rm Der}\nolimits}

\def\ker{\mathop{\rm ker}\nolimits}

\def\mone{{^{-1}}}
\def\demi{{\frac{1}{2}}}
\def\act{{\vartriangleright}}

\def\ff{{\cal F}}
\def\gg{{\cal G}}

\def\a2{{a^2}}

\let\tilde=\widetilde

\newcommand{\color}[2][c]{}

\numberwithin{equation}{section}

\makeatletter

\newfont{\@aidxte}{cmsy10}
\newfont{\@aidxel}{cmsy10 scaled 1095}
\newfont{\@aidxtw}{cmsy10 scaled 1200}
\newlength\@aidxtexvi
\newlength\@aidxtexvii
\newlength\@aidxelxvi
\newlength\@aidxelxvii
\newlength\@aidxtwxvi
\newlength\@aidxtwxvii
\newcommand{\alignidx}[1]{%
  \@aidxtexvi=\fontdimen16\@aidxte
  \@aidxtexvii=\fontdimen17\@aidxte
  \@aidxelxvi=\fontdimen16\@aidxel
  \@aidxelxvii=\fontdimen17\@aidxel
  \@aidxtwxvi=\fontdimen16\@aidxtw
  \@aidxtwxvii=\fontdimen17\@aidxtw
    {\mbox{$%
    \fontdimen16\@aidxte=2.9pt
    \fontdimen17\@aidxte=2.9pt
    \fontdimen16\@aidxel=3.1pt
    \fontdimen17\@aidxel=3.1pt
    \fontdimen16\@aidxtw=3.3pt
    \fontdimen17\@aidxtw=3.3pt
    #1$}}%
    \fontdimen16\@aidxte=\@aidxtexvi
    \fontdimen17\@aidxte=\@aidxtexvii
    \fontdimen16\@aidxel=\@aidxelxvi
    \fontdimen17\@aidxel=\@aidxelxvii
    \fontdimen16\@aidxtw=\@aidxtwxvi
    \fontdimen17\@aidxtw=\@aidxtwxvii}

\makeatother

\def\nn{\notag}
\def\del{\partial}
\def\emph#1{{\sl #1\/}}

\usepackage[ps,matrix,arrow,curve]{xy}
\xyoption{dvips} % if you switch this off, you need bigtex.

\begin{document}
%
% ==============================================================================

\preprint{DAMTP-2003-86}

\title{Higher gauge theory --- differential versus integral\\
       formulation}

\author{Florian Girelli}
\email{fgirelli@perimeterinstitute.ca}
\affiliation{Perimeter Institute for Theoretical Physics, 35 King Street N,
Waterloo, Ontario, N2J 2W9, Canada}

\author{Hendryk Pfeiffer}
\email{hpfeiffer@perimeterinstitute.ca}
\affiliation{Perimeter Institute for Theoretical Physics, 35 King Street N,
Waterloo, Ontario, N2J 2W9, Canada}
\affiliation{Emmanuel College, St Andrew's Street, Cambridge CB2 3AP, United
Kingdom}
\affiliation{DAMTP, Wilberforce Road, Cambridge CB3 0WA, United Kingdom}

\date{January 28, 2004}

% ==============================================================================
%
\begin{abstract}
%
% ==============================================================================

The term higher gauge theory refers to the generalization of gauge
theory to a theory of connections at two levels, essentially given by
$1$- and $2$-forms. So far, there have been two approaches to this
subject. The differential picture uses non-Abelian $1$- and $2$-forms
in order to generalize the connection $1$-form of a conventional gauge
theory to the next level. The integral picture makes use of curves and
surfaces labeled with elements of non-Abelian groups and generalizes
the formulation of gauge theory in terms of parallel transports. We
recall how to circumvent the classic no-go theorems in order to define
non-Abelian surface ordered products in the integral picture. We then
derive the differential picture from the integral formulation under
the assumption that the curve and surface labels depend smoothly on
the position of the curves and surfaces. We show that some aspects of
the no-go theorems are still present in the differential (but not in
the integral) picture. This implies a substantial structural
difference between non-perturbative and perturbative approaches to
higher gauge theory. We finally demonstrate that higher gauge theory
provides a geometrical explanation for the extended topological
symmetry of $BF$-theory in both pictures.
\end{abstract}


\maketitle

% ==============================================================================
%
\section{Introduction}
%
% ==============================================================================

Gauge theory can be formulated in two ways which we term the
differential and the integral picture. As an illustration, recall, for
example, Maxwell's equations which can be formulated either in terms
of integral equations relating electric and magnetic fluxes through
surfaces and currents through solenoids\footnote{This is actually the
form which corresponds to experimental setups and in which the laws of
electrodynamics were originally discovered.} (integral picture) or
alternatively in terms of the familiar differential equations
(differential picture).

Similarly, any gauge theory can be formulated in two ways. Let $M$ be
some space-time manifold and the gauge group $G$ be a Lie group with
Lie algebra $\g$. In the differential formulation of gauge theory, one
considers a connection of some principal $G$-bundle $P\to M$. In local
coordinates, \ie\ using a local trivialization of the bundle, the
connection is given by a $\g$-valued connection $1$-form $A$ which
transforms under changes of the coordinates as,
\begin{equation}
\label{eq_connection}
  A\mapsto A^\prime = g^{-1}Ag + g^{-1}dg,
\end{equation}
where $g\colon U_1\cap U_2\to G$ denotes a transition function on the
overlap of the coordinate patches $U_1,U_2\subseteq M$.

The connection is usually taken to be the basic field of the theory,
\ie\ the variation in the action principle is with respect to $A$, and
in a path integral quantum theory, one has to integrate over all
possible connections. The Lagrangian and the action of the theory
depend on the curvature,
\begin{equation}
\label{eq_curvature}
  F = dA + \frac{1}{2}[A,A],
\end{equation}
which is a $\g$-valued $2$-form. It transforms under coordinate
changes in a gauge covariant way,
\begin{equation}
  F\mapsto F^\prime = g^{-1}Fg.
\end{equation}

As an alternative to this differential picture, there exists the
integral formulation. In this formulation, one uses the group valued
parallel transports,
\begin{equation}
  U_\gamma = P\exp\bigl( \int_\gamma A\bigr)\in G,
\end{equation}
along curves $\gamma\colon[0,1]\to M, \tau\mapsto\gamma(\tau)$ as the
basic variables. Locally, the parallel transport always exists, and it
is uniquely determined as the solution of the first order matrix
differential equation,
\begin{equation}
  \frac{d}{dt}U_\gamma(t) = [A_\mu(\gamma(t))\dot\gamma^\mu(t)]U_\gamma(t),
\end{equation}
where we have written,
\begin{equation}
  U_\gamma(t) = P\exp\biggl( \int_0^t A_\mu(\gamma(\tau))\dot\gamma^\mu(\tau)d\tau\biggr),
\end{equation}
for the parallel transport along $\gamma$ from $\tau=0$ to
$\tau=t$. The initial condition is $U_\gamma(0)=1\in G$.

The curvature can then be calculated from the holonomy $U_\gamma$ of a
closed loop $\gamma$ in the limit in which the loop shrinks to
infinitesimal size. This integral picture of gauge theory closely
resembles what is done in lattice gauge theory, but without the
restriction that the curves have to live on some fixed lattice. Note
that in the integral formulation, the parallel transports satisfy
various relations.

Now let us illustrate the basic idea of higher gauge theory. Suppose
first that the gauge group in conventional gauge theory is Abelian,
say $G=\U(1)$. Then the connection $1$-form $A$ is imaginary and the
transition functions are of the form $g(x)=e^{\phi(x)}$, where
$\phi\colon U\to i\R$ is a suitable imaginary valued function,
and~\eqref{eq_connection} becomes,
\begin{equation}
\label{eq_abelconnection}
  A\mapsto A^\prime = A + d\phi,
\end{equation}
so that the curvature $2$-form~\eqref{eq_curvature} is just the
exterior derivative,
\begin{equation}
\label{eq_abelcurvature}
  F = dA.
\end{equation}

Abelian gauge theory with~\eqref{eq_abelconnection}
and~\eqref{eq_abelcurvature} now admits the following higher level
generalization. Let $A$ be some (imaginary valued) $p$-form which
becomes the basic field of the theory. The Lagrangian and the action
depend only on the curvature $(p+1)$-form $F=dA$. The theory therefore
enjoys a local gauge symmetry with the transformation
law~\eqref{eq_abelconnection} for some $(p-1)$-form $\phi$. This is a
consequence of the Poincar\'e lemma because locally any closed
$p$-form $A-A^\prime$ is of the form $d\phi$. The Abelian theory at
level $p$ is therefore completely governed by the de~Rham cohomology
of $M$.

The Abelian theory at level $p=2$ is known in the physics literature
as Kalb--Ramond fields~\cite{KaRa74}, and at generic level $p$ as
$p$-form electrodynamics~\cite{HeTe86}. Both refer to the differential
picture of the theory. The corresponding integral picture makes use of
$p$-dimensional surfaces labeled with elements of the Abelian group
$\U(1)$. If the $p$-surfaces were restricted to a fixed hyper-cubic
lattice, we would have at level $p=0$ the $xy$-model of statistical
mechanics, at $p=1$, $\U(1)$-lattice gauge theory, and at higher $p$
the models of~\cite{Sa77}.

We refer to these higher level theories as \emph{higher gauge
theory}. Does there exist a \emph{non-Abelian} higher gauge theory, at
least at level $p=2$? Many
authors~\cite{Ne83,Or83,So93,La02a,La02b,Ch02,Ho02} have attempted to
construct such models, but the necessity to find an underlying
geometrical picture by suitably generalizing fibre bundles seems to
impose serious constraints. To our knowledge, the only thoroughly
studied model was the Freedman--Townsend model~\cite{FrTo81} which,
however, has only an Abelian local symmetry and lacks a geometrical
understanding of why it has to be Abelian.

What precisely are the geometrical conditions involved in higher gauge
theory? The coboundary condition $d\circ d=0$ of the Abelian case is,
of course, no longer useful as the curvature $F$ is in general no
longer just $dA$, but rather given by~\eqref{eq_curvature}. Assume we
have some non-Abelian `connection' $2$-form and wish to define its
`curvature' $3$-form. Geometrically, the $2$-form would be associated
with surfaces labeled by elements of some non-Abelian group $H$, and
the curvature $3$-form should be some group element associated with a
closed surface, composed from several constituent surfaces that are
all labeled with elements of $H$. Since the group $H$ is non-Abelian
and there is no canonical surface order available, we simply do not
know how to compose the various non-Abelian labels.

Recall that the points of a curve have a natural order, and the
definition of the parallel transport along a given curve indeed makes
use of this order. For higher dimensional submanifolds, however, such
a canonical order is not available. This lack of natural order has led
Teitelboim~\cite{Te86} to the formulation of a no-go theorem ruling
out the existence of non-Abelian gauge theories for extended objects.
This applies essentially to any gauge theory whose connection is a
non-Abelian $p$-form, $p\geq 2$.

With the introduction of $2$-categories in mathematics, it recently
became possible to sidestep this no-go theorem at $p=2$. On the
mathematical side, there is the construction of non-Abelian gerbes
generalizing fibre bundles, see, for example~\cite{BrMe01}. It is
expected that gerbes provide the desired generalization of fibre
bundles. They are, for example, conjectured to play a role in theories
on coincident $5$-branes is string theory, see, for
example~\cite{St96}. In the present paper, we prefer a slightly
different approach based on the definition of Lie $2$-groups by
Baez~\cite{Ba02} which generalize ordinary Lie groups to higher level
and which include the symmetries of gerbes at least in the case of
strict categories. This will allow us to explicitly solve the surface
ordering problem, thereby providing a rigorous basis for Chepelev's
conjectures~\cite{Ch02}, and to see in detail how Teitelboim's no-go
theorem is avoided. Our results finally provide the geometrical
background to most of the traditional approaches to non-Abelian
$2$-forms, at least as long as strict categories are sufficient, and
explain geometrically why these models are so restricted. It comes as
a surprise that $BF$-theory which is usually not mentioned in the
context of non-Abelian $2$-forms, does form a non-trivial example of
higher gauge theory.

Let us now briefly outline our approach. Starting from the notion of
Lie $2$-groups, Baez has defined Lie $2$-algebras and started to
generalize the differential picture of gauge theory to a theory
involving non-Abelian connection $1$- and $2$-forms~\cite{Ba02}. Open
questions in this approach are the precise form of the local gauge
transformations and of the gauge invariant expressions which are
required in order to define Lagrangians and actions in physics.

Also starting from the notion of Lie $2$-groups, we have generalized
the integral picture of gauge theory to a theory involving curves and
surfaces labeled with elements of non-Abelian groups~\cite{Pf03}. This
formulation has the advantage that the theory of $2$-categories
dictates the form of the local gauge transformations and the
expressions for the gauge invariant quantities. The no-go theorems can
be avoided because the underlying $2$-categorical structure leads to a
non-trivial interplay of the curve and surface labels.

An important question is how the differential~\cite{Ba02} and the
integral approach~\cite{Pf03} are related. In this article, we start
from the integral picture of~\cite{Pf03} and systematically derive the
corresponding differential expressions by studying the non-Abelian
curve and surface labels of the theory in the infinitesimal limit,
assuming that the labels depend smoothly on the curves and
surfaces. In the smooth case, we find an additional flatness condition
at level $1$ which has not yet appeared in the literature. It implies
in particular that the non-Abelian part of the connection $2$-form
agrees with the curvature of the connection $1$-form, that the
curvature $2$-form vanishes and that the curvature $3$-form is
Abelian.

We show that an interesting example of higher gauge theory is given by
$BF$-theory with non-Abelian gauge group~\cite{Ho89} in which the
level-$1$ flatness is a key feature of the theory, in fact encoded in
the field equations. The theory of $2$-categories then provides the
explanation for the extended local (topological) symmetry. Otherwise,
the resulting conditions show that the classic no-go theorems reappear
only in the differential picture in which they rule out the naive
generalization of the Yang--Mills action. The algebraic structure of
the integral picture, however, still allows us to have non-trivial
central group elements that characterize the $2$-curvature associated
with closed labeled surfaces which we call $2$-holonomies. If the
centre of the gauge group is discrete such as, for example, for
$\SU(N)$, the differential picture would require these central
elements to be trivial, but in the integral picture and in any
non-perturbative quantum theory based on it, no such restriction
applies. The integral picture is therefore more general than the
differential one and is in some sense essentially non-perturbative.

Since the non-trivial central elements can be interpreted as the
presence of singularities of codimension $2$, in some sense the
integral picture of higher gauge theory rather \emph{predicts} the
existence of topological defects in the differential formulation.

The present article is structured as follows. In
Section~\ref{sect_integral}, we recall the construction of higher
gauge theory in the integral formulation as it was developed
in~\cite{Pf03}. Our presentation is self-contained. We emphasize the
calculational aspects and try to hide as much as possible of the
$2$-category theory in our notation. In
Section~\ref{sect_differential}, we then derive the differential
formulation of higher gauge theory starting from the integral picture
and compare the result with~\cite{Ba02}. We conclude in
Section~\ref{sect_conclusion} with comments on interesting questions
for further investigations.

% ==============================================================================
%
\section{The integral formulation}
%
% ==============================================================================
\label{sect_integral}

In this section, we review the integral picture of higher gauge theory
following~\cite{Pf03}. We call the higher level model a
\emph{$2$-gauge theory} whereas we refer to conventional gauge theory
as a \emph{$1$-gauge theory} in view of the hierarchy of models
sketched in~\cite{Pf03}. The theory is formulated at the integral
level, \ie\ it describes curves and surfaces which are labeled with
data from some algebraic structure, supplementing the parallel
transports of conventional $1$-gauge theory by additional group
elements which are used to label surfaces.

%------------------------------------------------------------------------------
\subsection{Lie $2$-groups}
%------------------------------------------------------------------------------
\label{sect_2group}

The algebraic structure required to describe a $1$-gauge theory is
just some gauge group $G$, usually taken to be a Lie group. The
geometric objects that are labeled with algebraic data, are the curves
giving rise to the parallel transports of the theory. The group
structure ensures that we can consistently compose (multiply) parallel
transports and also reverse their direction (inversion).

The algebraic structure for a $2$-gauge theory is a so-called
$2$-group~\cite{Ba02,Fo02}. This is a pair $G$, $H$ of groups with two
maps\footnote{We define here a \emph{crossed module}, a structure from
which we can construct a strict $2$-group~\cite{Ba02,Fo02}.}. The
first map is a group homomorphism $t\colon H\to G$, \ie
\begin{eqnarray}
  t(h_1\cdot h_2)&=&t(h_1)\cdot t(h_2),\\
  t(1)&=&1,
\end{eqnarray}
for all $h_1,h_2\in H$. The second map is an action of $G$ on $H$ by
automorphisms. This is an operation $g\rhd h$ taking values in $H$,
which is a group action, \ie
\begin{eqnarray}
  (g_1\cdot g_2)\rhd h &=& g_1\rhd (g_2\rhd h),\\
  1\rhd h &=& h,
\end{eqnarray}
for all $g_1,g_2\in G$ and $h\in H$, such that $h\mapsto g\rhd h$ is a
homomorphism for each $g\in G$, \ie\
\begin{eqnarray}
  g\rhd (h_1\cdot h_2)&=&(g\rhd h_1)\cdot(g\rhd h_2),\\
  g\rhd 1 &=& 1,
\end{eqnarray}
for all $h_1,h_2\in H$. These maps are required to satisfy the
following two compatibility conditions,
\begin{eqnarray}
\label{eq_compatibility}
  t(g\rhd h) &=& g\cdot t(h)\cdot g^{-1},\\
  t(h)\rhd h^\prime &=& h\cdot h^\prime\cdot h^{-1}.
\end{eqnarray}
for all $g\in G$, $h,h^\prime\in H$. A Lie $2$-group is a $2$-group in
which $G$ and $H$ are Lie groups and both maps $t$ and $\rhd$ are
smooth.

Plenty of examples of Lie $2$-groups are known~\cite{Ba02,Pf03}. Here
we mention the following cases.
\begin{enumerate}
\item
  The \emph{Euclidean} and \emph{Poincar\'e $2$-groups}. Here $H=\R^n$
  is Euclidean or Minkowski space and $G=\SO(n)$ or $\SO(n-1,1)$. The
  map $t$ is trivial, \ie\ $t(h)=1$ for all $h\in H$, and $\rhd$ is
  the obvious action by rotation.
\item
  More generally, one can choose $H$ to be any vector space on which
  the Lie group $G$ is represented. The map $t$ is trivial in this
  case, and $\rhd$ is the action of $G$ on its representation $H$. In
  this way, one defines, for example, the \emph{adjoint} and the
  \emph{co-adjoint $2$-groups} in which $H=\g$ or $H=\g^*$ where $\g$
  denotes the Lie algebra of $G$.
\item
  The \emph{automorphism $2$-groups}. Let $H$ be any Lie group and $G$
  its group of automorphisms. The action $g\rhd h$ is the application
  of the particular automorphism, and $t\colon H\to G$ associates with
  each element $h$ the corresponding \emph{inner automorphism}
  $h^\prime\mapsto hh^\prime h^{-1}$. This example is related to
  non-Abelian gerbes~\cite{Ba02}. For example, if $H=\SU(2)$, we have
  $G=SU(2)/\Z_2\cong SO(3)$ where $\Z_2$ is the centre of $\SU(2)$. In
  general, $\ker t\subseteq Z(H)$ is always contained in the centre of
  $H$.
\end{enumerate}

%------------------------------------------------------------------------------
\subsection{$2$-gauge theory}
%------------------------------------------------------------------------------
\label{sect_2gauge}

In a $2$-gauge theory, we have to label geometric objects at two
levels. Curves are labeled by elements of $G$. Their composition and
orientation reversal is defined as in conventional gauge theory.

In addition, surfaces are labeled with elements of $H$. For each
surface\footnote{The elementary surfaces are chosen to have the
topology of a disc.}, we choose two reference points on the boundary
(full dots in the diagram below, we are going to suppress them later
on) and split the boundary into two curves with labels $g_1\in G$
(\emph{source}) and $g_2\in G$ (\emph{target}) as follows,
\begin{equation}
\label{eq_surface}
\begin{aligned}
\xymatrix{
  \bullet\ar@/^2ex/[rr]^{g_1}="g1"\ar@/_2ex/[rr]_{g_2}="g2"&&\bullet
  \ar@{=>}^h "g1"+<0ex,-2ex>;"g2"+<0ex,2ex>
}
\end{aligned}
\end{equation}
The label $h$ of the surface is required to satisfy,
\begin{equation}
\label{eq_constraint}
  t(h)=\alignidx{g_2\cdot g_1^{-1}},
\end{equation}
\ie\ $t(h)$ is the (inverse) holonomy along the boundary curve.  This
condition appears when we use the Lie crossed module in order to
construct a Lie $2$-group~\cite{Ba02,Pf03,Fo02}. The first reference
point is the base point of this holonomy and therefore plays a role
in~\eqref{eq_constraint} whereas the second reference point does not
enter this condition.

We can now compose surfaces in two different ways. Firstly, we can
join them \emph{horizontally} in one common reference point,
\begin{equation}
\label{eq_horizontal}
\begin{aligned}
\xymatrix{
  \ar@/^2ex/[rr]^{g_1}="g1"\ar@/_2ex/[rr]_{g_2}="g2"&&
  \ar@/^2ex/[rr]^{g_1^\prime}="g3"\ar@/_2ex/[rr]_{g_2^\prime}="g4"&&
  \ar@{=>}^{h} "g1"+<0ex,-2ex>;"g2"+<0ex,2ex>
  \ar@{=>}^{h^\prime} "g3"+<0ex,-2.5ex>;"g4"+<0ex,2.5ex>
}
=
\xymatrix{
  \ar@/^2ex/[rr]^{g_1\cdot g_1^\prime}="g1"\ar@/_2ex/[rr]_{g_2\cdot g_2^\prime}="g2"&&
  \ar@{=>}^{\tilde h} "g1"+<0ex,-2.5ex>;"g2"+<0ex,2.5ex>
}
\end{aligned}
\end{equation}
where the label of the composition is given by
\begin{equation}
\label{eq_horizontaleq}
  \tilde h=h\cdot (g_1\rhd h^\prime).
\end{equation}
Note the asymmetry: the source of the first surface acts on the label
of the second one\footnote{The pairs $(h,g_1)$ of surface label and
source curve label form the semi-direct product $H\rtimes G$ under
horizontal composition.}. As required, it follows that $t(\tilde
h)=\alignidx{(g_2g_2^\prime){(g_1g_1^\prime)}^{-1}}$.  Alternatively,
we can glue the surfaces \emph{vertically} along a common curve,
\begin{equation}
\label{eq_vertical}
\begin{aligned}
\xymatrix{
  \ar@/^4ex/[rr]^{g_1}="g1"\ar[rr]^(0.35){g_2}\ar@{}[rr]|{}="g2"
  \ar@/_4ex/[rr]_{g_3}="g3"&&
  \ar@{=>}^h "g1"+<0ex,-2ex>;"g2"+<0ex,1ex>
  \ar@{=>}^{h^\prime} "g2"+<0ex,-1ex>;"g3"+<0ex,2ex>
}
=
\xymatrix{
  \ar@/^2ex/[rr]^{g_1}="g1"\ar@/_2ex/[rr]_{g_3}="g2"&&
  \ar@{=>}^{\tilde h} "g1"+<0ex,-2ex>;"g2"+<0ex,2ex>
}
\end{aligned}
\end{equation}
where the composition is simply given by
\begin{equation}
\label{eq_verticaleq}
  \tilde h=h^\prime\cdot h.
\end{equation}
Observe that $t(\tilde h)=\alignidx{g_3g_1^{-1}}$ as expected.

The orientation of a surface can be reversed if it is labeled by the
inverse element $h^{-1}$ instead,
\begin{equation}
\label{eq_orient1}
\begin{aligned}
\xymatrix{
  \ar@/^2ex/[rr]^{g_1}="g1"\ar@/_2ex/[rr]_{g_2}="g2"&&
  \ar@{=>}^h "g1"+<0ex,-2ex>;"g2"+<0ex,2ex>
}
=
\xymatrix{
  \ar@/^2ex/[rr]^{g_1}="g1"\ar@/_2ex/[rr]_{g_2}="g2"&&
  \ar@{=>}_{h^{-1}} "g2"+<0ex,2ex>;"g1"+<0ex,-2ex>
}
\end{aligned}
\end{equation}
Both source and target curve of some surface can be reversed,
\begin{equation}
\label{eq_orient2}
\begin{aligned}
\xymatrix{
  \ar@/^2ex/[rr]^{g_1}="g1"\ar@/_2ex/[rr]_{g_2}="g2"&&
  \ar@{=>}^h "g1"+<0ex,-2ex>;"g2"+<0ex,2ex>
}
=
\xymatrix{
  &&\ar@/_2ex/[ll]_{g_1^{-1}}="g1"\ar@/^2ex/[ll]^{g_2^{-1}}="g2"
  \ar@{=>}^{\tilde h} "g1"+<0ex,-2.5ex>;"g2"+<0ex,2.5ex>
}
\end{aligned}
\end{equation}
if the surface label is replaced by $\tilde h=g_1^{-1}\rhd h^{-1}$.
Observe that $t(\tilde h)=\alignidx{g_2^{-1}{(g_1^{-1})}^{-1}}$ as
required.

An important operation is known as \emph{whiskering}. By attaching
whiskers to a surface $h$, for example attaching whiskers $g_1$ and
$g_1^{\prime\prime}$ to some surface $h^\prime$ with source
$g_1^\prime$ and target $g_2^\prime$,
\begin{equation}
\label{eq_whisker}
\begin{aligned}
\xymatrix{
  \ar[r]^{g_1}&\ar@/^2ex/[rr]^{g_1^\prime}="g1"\ar@/_2ex/[rr]_{g_2^\prime}="g2"&&
  \ar[r]^{g_1^{\prime\prime}}&
  \ar@{=>}^{h^\prime} "g1"+<0ex,-2.5ex>;"g2"+<0ex,2.5ex>
}
=
\xymatrix{
  \ar@/^2ex/[rr]^{g_1g_1^\prime g_1^{\prime\prime}}="g1"\ar@/_2ex/[rr]_{g_1g_2^\prime g_1^{\prime\prime}}="g2"&&
  \ar@{=>}^{\tilde h^\prime} "g1"+<0ex,-2.5ex>;"g2"+<0ex,2.5ex>
}
\end{aligned}
\end{equation}
we can construct a surface with source $\alignidx{g_1g_1^\prime
g_1^{\prime\prime}}$ and target $\alignidx{g_1g_2^\prime
g_1^{\prime\prime}}$, carrying the label
\begin{equation}
\label{eq_whiskereq}
\tilde h^\prime=g_1\rhd h^\prime.
\end{equation}
The attachment of the left whisker can be understood as a special case
of the horizontal composition~\eqref{eq_horizontal} in which $g_1=g_2$
and $h=1$ so that the left surface collapses to a line. A similar
argument is available for the right whisker. The asymmetry in the
expression~\eqref{eq_whiskereq} originates from the asymmetry of the
horizontal composition~\eqref{eq_horizontaleq}.

Whiskering allows us to change the reference points of a surface. For
example, starting from a surface $h$ with reference points $x$ and
$y$, \ie\ source $g_1\cdot g_2$ and target $g_3$,
\begin{equation}
\begin{aligned}
\xymatrix{
  &x\ar@<.5ex>[dl]^{g_1}="g1"\ar[dr]^{g_3}="g3"\\
  z\ar@<.5ex>[ur]^{g_1^{-1}}\ar[rr]_{g_2}&&
  y\ar@{=>}_h "g1"+<3ex,0ex>;"g3"+<-3ex,-1ex>
}
\end{aligned}
=
\begin{aligned}
\xymatrix{
  &x\ar[dr]^{g_3}\\
  z\ar[ur]^{g_1^{-1}}\ar[rr]_{g_2}="g2"&&
  y\ar@{=>}_{\tilde h} "g2"+<0ex,3ex>;"1,2"+<0ex,-3ex>
}
\end{aligned}
\end{equation}
we can whisker from the left by $g_1^{-1}$ and obtain the surface
$\tilde h=g_1^{-1}\rhd h$ with reference points $z$ and $y$, \ie\
source $g_2$ and target $g_1^{-1}\cdot g_3$.

Given any collection of curves and surfaces, a \emph{configuration} of
$2$-gauge theory is an assignment of elements of $G$ to the curves and
of elements of $H$ to the surfaces so that the following conditions
hold. Compositions of curves are labeled by the product of elements in
$G$, curves of opposite orientation are labeled by the inverse group
element. For each surface labeled by $h\in H$, we have
$t(h)=\alignidx{g_2\cdot g_1^{-1}}$ where $g_1$ and $g_2$ are the
source and target curve, respectively. Finally, compositions of
surfaces, and surfaces whose reference points have been changed, are
labeled as described above in this section. The configurations thus
defined can be viewed as the classical configurations of $2$-gauge
theory or, in a path integral quantum theory, these are the
configurations over which we sum in the path integral. The path
integral was given in detail in~\cite{Pf03}.

%------------------------------------------------------------------------------
\subsection{Local $2$-gauge transformations}
%------------------------------------------------------------------------------

The $2$-gauge theory defined in the preceding section enjoys an
extended local gauge symmetry which we call a \emph{local $2$-gauge
symmetry}.

First recall the conventional local $1$-gauge symmetry in a
formulation of gauge theory in the language of parallel transports. A
local gauge transformation is given by a \emph{generating function}
assigning group elements $\eta_x,\eta_y\in G$ to the points. For each
curve $\gamma$ from point $x$ to point $y$ with label $g_\gamma\in G$,
the transformed parallel transport is then calculated by,
\begin{equation}
\label{eq_1gauge}
  \alignidx{{\tilde g}_\gamma = \eta_x^{-1}g_\gamma\eta_y},
\end{equation}
which we visualize by the following diagram,
\begin{equation}
\label{eq_1gaugediag}
\begin{aligned}
\xymatrix{
  x\ar[rr]^{g_\gamma}\ar[dd]_{\eta_x}&&y\ar[dd]^{\eta_y}\\
  \\
  x\ar[rr]_{{\tilde g}_\gamma}&&y
}
\end{aligned}
\end{equation}
We say that the diagram \emph{commutes}, \ie\ it does not matter which
way round we go from one corner to another. If we view all four
labeled curves $g_\gamma$, ${\tilde g}_\gamma$, $\eta_x$ and $\eta_y$
as a gauge connection, then this connection is \emph{flat}, \ie\ the
parallel transport is path independent.

In $2$-gauge theory, the local gauge transformation~\eqref{eq_1gauge}
is weakened by extending the generating function to the next
level. The \emph{$2$-generating function} not only assigns group
elements $\eta_x,\eta_y\in G$ to the points, but there is the
additional freedom of choosing elements $\eta_\gamma\in H$ for the
curves\footnote{These curve labels are in general path
dependent.}. Diagram~\eqref{eq_1gaugediag} is generalized to
\begin{equation}
\begin{aligned}
\label{eq_2gauge1diag}
\xymatrix{
  x\ar[rr]^{g_\gamma}\ar[dd]_{\eta_x}&&y\ar[dd]^{\eta_y}\\
  \\
  x\ar[rr]_{{\tilde g}_\gamma}&&y
  \ar@{=>}_{\eta_\gamma} "1,3"+<-3ex,-3ex>; "3,1"+<6ex,6ex>
}
\end{aligned}
\end{equation}
where we require $t(\eta_\gamma)=\alignidx{(\eta_x{\tilde
g}_\gamma){(g_\gamma\eta_y)}^{-1}}$. The full diagram involving
$g_\gamma$, ${\tilde g}_\gamma$, $\eta_x$, $\eta_y$ and $\eta_\gamma$
can therefore be viewed as a configuration of $2$-gauge theory in
which the surface labeled with $\eta_\gamma$ has the source
$g_\gamma\cdot\eta_y$ and the target $\eta_x\cdot{\tilde
g}_\gamma$. We can thus calculate the gauge transformed parallel
transport by,
\begin{equation}
\label{eq_2gauge1}
  \alignidx{{\tilde g}_\gamma = \eta_x^{-1}t(\eta_\gamma)g_\gamma\eta_y},
\end{equation}
which generalizes the conventional local gauge
transformation~\eqref{eq_1gauge}.

This is the prescription of how to transform the curve labels. In
$2$-gauge theory, we have to specify in addition how to transform the
surface labels. Therefore we write down the surface analogue of the
diagram~\eqref{eq_1gaugediag} and require that for each surface
labeled $h\in H$ with source and target curves $\gamma$,
$\gamma^\prime$ labeled by $g_\gamma$ and $g_{\gamma^\prime}$, the
following `tin can' diagram
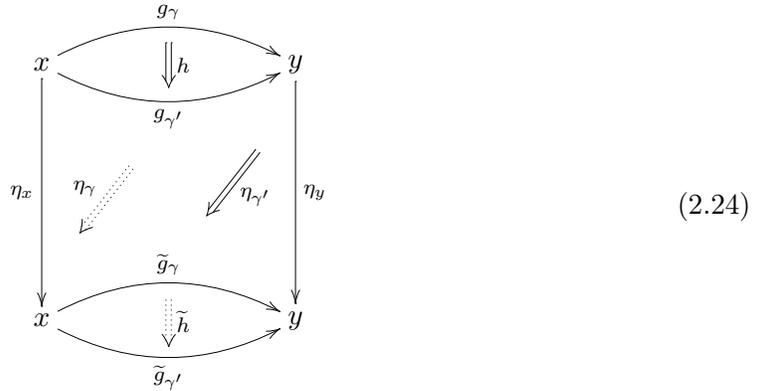
\begin{equation}
\label{eq_tincan}
\begin{aligned}
\xymatrix{
  x\ar[ddd]_{\eta_x}
    \ar@{}[ddd]^(0.85){}="fx"\ar@/^3ex/[rrr]^{g_\gamma}="a"\ar@/_3ex/[rrr]_{g_{\gamma^\prime}}="b"
  &&&y\ar[ddd]^{\eta_y}\ar@{}[ddd]_(0.15){}="fy"\\
  \\
  \\
  x\ar@/^3ex/[rrr]^{{\tilde g}_\gamma}="c"\ar@/_3ex/[rrr]_{{\tilde g}_{\gamma^\prime}}="d"&&&y\\
  \ar@{=>}^{h} "a"+<0pt,-2.5ex>;"b"+<0pt,2.5ex>
  \ar@{:>}^{\tilde h} "c"+<0pt,-2.5ex>;"d"+<0pt,2.5ex>
  \ar@{} "fy";"c"|(0.3){}="f1"
  \ar@{} "fy";"c"|(0.7){}="f2"
  \ar@{} "b";"fx"|(0.3){}="b1"
  \ar@{} "b";"fx"|(0.7){}="b2"
  \ar@{=>} "f1";"f2"^{\eta_{\gamma^\prime}}
  \ar@{:>} "b1";"b2"_{\eta_\gamma}
}
\end{aligned}
\end{equation}
\emph{$2$-commutes}. This means that it does not matter which way
round we compose the labeled surfaces, \ie\ the configuration of
$2$-gauge theory shown in diagram~\eqref{eq_tincan} is
\emph{$2$-flat}. The top of this `tin can' is the old configuration
and the bottom the new one with curve labels ${\tilde g}_\gamma$,
${\tilde g}_{\gamma^\prime}$ and surface label $\tilde h$. The
transformed surface label is thus given by,
\begin{equation}
\label{eq_2gauge2}
  {\tilde h} = \eta_x^{-1}\rhd\bigl(\eta_{\gamma^\prime} h\eta_\gamma^{-1}\bigr).
\end{equation}

We can summarize this paragraph as follows. The local $2$-gauge
transformations are given by $2$-generating functions which assign
elements of $G$ to the points and elements of $H$ to the curves. The
transformed curve and surface labels are then determined
by~\eqref{eq_2gauge1} and~\eqref{eq_2gauge2}. Although, at first
sight, these transformation rules look quite artificial, they follow
immediately from the underlying $2$-categorical structure~\cite{Pf03}.

%------------------------------------------------------------------------------
\subsection{Pure $2$-gauge and $2$-flatness}
%------------------------------------------------------------------------------
\label{sect_puregauge}

In conventional $1$-gauge theory, we say that a configuration is
\emph{pure gauge} if it is gauge equivalent to the trivial connection
in which all curves are assigned the group unit. A configuration is
therefore pure gauge if there exists a generating function associating
group elements $\eta_j\in G$ with all points so that the parallel
transports are given by
\begin{equation}
  \alignidx{g_{12} = \eta_1^{-1}\eta_2},
\end{equation}
for any curve from $1$ to $2$, \cf~\eqref{eq_1gauge}. Observe that any
configuration which is pure gauge, is also \emph{flat}, \ie\ its
parallel transports are path independent.

In complete analogy, we say that a configuration of $2$-gauge theory
is \emph{pure $2$-gauge} if it is $2$-gauge equivalent to the trivial
configuration in which all curves are labeled by the group unit of $G$
and all surfaces by the group unit of $H$. A configuration with curve
labels $g_\gamma,g_{\gamma^\prime}\in G$ and surface labels $h\in H$
is therefore pure $2$-gauge if there exists a $2$-generating function
that assigns elements $\eta_x,\eta_y\in G$ to the points and
$\eta_\gamma,\eta_{\gamma^\prime}\in H$ to the curves such that for
any curve $\gamma$ from $x$ to $y$,
\begin{equation}
\label{eq_gflat}
  g_\gamma = \alignidx{\eta_x^{-1}t(\eta_\gamma)\eta_y},
\end{equation}
and for any surface with source curve $g_\gamma$ and target curve
$g_{\gamma^\prime}$,
\begin{equation}\label{eq_hflat}
  h=\eta_x^{-1}\rhd (\alignidx{\eta_{\gamma^\prime}\eta_\gamma^{-1}}),
\end{equation}
\cf~\eqref{eq_2gauge1} and~\eqref{eq_2gauge2}.

A configuration of $2$-gauge theory is called \emph{$2$-flat} if the
surface label on any surface of topology $S^2$ which is the boundary
of a $3$-ball, is just the group unit $1\in H$. As a consequence, in
$2$-flat configurations, the surface label of any disc shaped surface
depends only on the boundary (source and target) curve labels in
$G$. Note also that being pure $2$-gauge implies $2$-flatness.

%------------------------------------------------------------------------------
\subsection{Composition of labeled surfaces}
%------------------------------------------------------------------------------

In this section, we describe how the language of $2$-gauge theory can
be used in order to define compositions of labeled surfaces. We will
make use of this surface composition in Section~\ref{sect_2gaugeinv}
in order to construct gauge invariant quantities that are associated
with closed surfaces, and in Section~\ref{sect_differential} in order
to derive the differential formulation.

In Section~\ref{sect_2gauge}, we have introduced a number of
operations by which we can modify and combine labeled surfaces:
vertical and horizontal composition, two types of orientation reversal
and the change of reference point by whiskering. These rules can be
employed in order to calculate the composition of elementary surfaces
to arbitrarily large ones.

We illustrate this procedure for the boundary surface of a
tetrahedron,
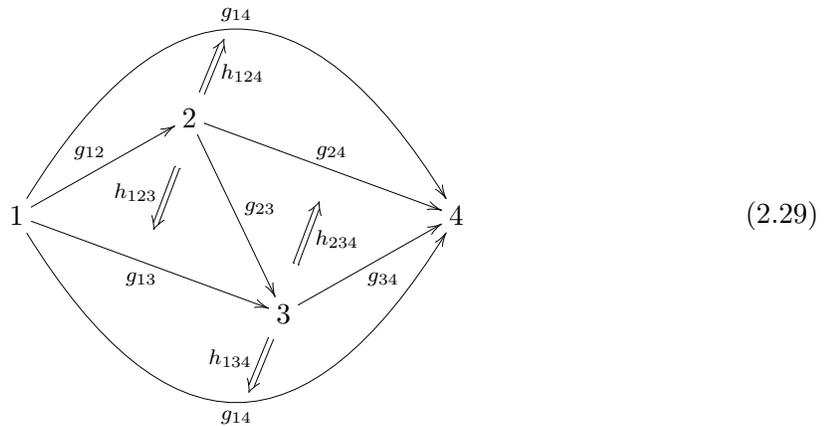
\begin{equation}
\label{eq_tetrahedron}
\begin{aligned}
\xymatrix{
  &&2\ar[drrr]^{g_{24}}="d"\ar[ddr]^{g_{23}}\\
  1\ar@/^15ex/[rrrrr]^{g_{14}}="a"\ar@/_15ex/[rrrrr]_{g_{14}}="b"\ar[drrr]_{g_{13}}="c"\ar[urr]^{g_{12}}&&&&&4\\
  &&&3\ar[urr]_{g_{34}}
  \ar@{=>}_{h_{124}} "1,3"+<1ex,2ex>;"a"+<-1ex,-2ex>
  \ar@{=>}_{h_{134}} "3,4"+<-1ex,-2ex>;"b"+<1ex,2ex>
  \ar@{=>}_{h_{123}} "1,3"+<-1ex,-4ex>;"c"+<1ex,4ex>
  \ar@{=>}_{h_{234}} "3,4"+<1ex,4ex>;"d"+<-1ex,-4ex>
}
\end{aligned}
\end{equation}
We have numbered the vertices by $1,2,3,4$. The edges $(j,k)$, $j<k$,
are labeled by group elements $g_{jk}\in G$ and the triangles
$(j,k,\ell)$, $j<k<\ell$, by elements $h_{jk\ell}\in H$. We have
oriented the triangles $(j,k,\ell)$ so that they have the source
$g_{jk}\cdot g_{k\ell}$ and the target $g_{j\ell}$, \ie\
$t(h_{jk\ell})=\alignidx{g_{j\ell}{(g_{jk}g_{k\ell})}^{-1}}$.

We choose reference points, here $1$ and $4$, and cut the tetrahedron
surface along the edge $(14)$. This \emph{base edge} forms both the
source and the target curve of the surface. Imagine that a curve
starting from the source sweeps out the entire surface until it
reaches the target. This determines the ordering of the vertical
composition of the constituent surfaces. We just have to make sure
that all surfaces are composable, \ie\ they have the suitable
reference points and the correct orientation in order to compose them
vertically by~\eqref{eq_vertical}.

Consider the diagram~\eqref{eq_tetrahedron}. We first move the curve
from $g_{14}$ to $g_{12}g_{24}$ via $h_{124}^{-1}$. At this stage we
cannot compose the result with the triangle $(123)$ because source and
target would not match, but we can use the orientation reversed
triangle $(234)$, whiskered from the left by $g_{12}$. This moves our
curve to $g_{12}g_{23}g_{34}$ using the label $g_{12}\rhd
h_{234}^{-1}$ of the whiskered and reversed surface. In the next step,
we can use the triangle $(123)$, whiskered from the right by $g_{34}$
which does not change the label $h_{123}$. Finally, we move our curve
from $g_{13}g_{34}$ to $g_{14}$ along $h_{134}$.

The label associated to the boundary surface of the tetrahedron is
therefore the vertical composition, \cf~\eqref{eq_vertical},
\begin{equation}
\label{eq_tetcomposite}
  \tilde h = h_{134}h_{123}(g_{12}\rhd h_{234}^{-1})h_{124}^{-1}.
\end{equation}
This is a useful notation for the automorphism $2$-group in which
typically both $G$ and $H$ are non-Abelian. In the case of the
Euclidean and the Poincar\'e $2$-groups, it is preferable to write the
group structure of $H$ additively, \ie\
\begin{equation}
  \tilde h = h_{134} + h_{123} - g_{12}\rhd h_{234} - h_{124}.
\end{equation}

The following geometrical picture illustrates the surface
composition. Imagine the surface labels $h_{jk\ell}$ are interpreted
in a local coordinate system associated with their first reference
point $j$, the common starting point of their source and target
curves. If we vertically compose surfaces that are based at the same
reference point, \ie\ whose labels are given in the same coordinate
system, the composition is just the group product in $H$,
\cf~\eqref{eq_verticaleq}. If the reference points and therefore the
coordinate systems are different, however, then we have to parallel
transport before we can compare and multiply their surface labels. In
the example~\eqref{eq_tetcomposite}, this is relevant for the surface
$h_{234}$, the only surface that is not based at point $1$ but rather
at $2$. We have to whisker $h_{234}^{-1}$ from the left by $g_{12}$ in
order to obtain a surface $g_{12}\rhd h_{234}^{-1}$ with reference
point $1$.

For a closed surface of topology $S^2$, \ie\ of genus zero, source and
target curve agree so that $t(\tilde h)=1$. Recall that $\ker
t\subseteq Z(H)$ is always contained in the centre $Z(H)$ of $H$ and
therefore Abelian. We call the labels $h\in\ker t$ of closed surfaces
the \emph{$2$-holonomies} of the theory.

%------------------------------------------------------------------------------
\subsection{Gauge invariant expressions}
%------------------------------------------------------------------------------
\label{sect_2gaugeinv}

For all the assignments of algebraic data to geometric objects, we
should understand how they depend on the choices made. Consider, for
example, the holonomy along a closed loop (\emph{Wilson loop}) in
conventional gauge theory. It still depends on the base point of the
loop. It does so, however, in a well-understood way. Changing the base
point leads to the conjugation of the holonomy with the parallel
transport from the old to the new base point. Any group character
applied to the holonomy yields an invariant. Observe that the
independence of the base point and the invariance under local gauge
transformations are both implemented by the same operation, namely by
calculating the character. Due to its gauge invariance, the character
can then serve as the Lagrangian or as the action of a physical
theory.

An analogous result can be shown for the integral picture of $2$-gauge
theory~\cite{Pf03}. Consider a closed surface of topology $S^2$, for
example the surface of the tetrahedron~\eqref{eq_tetrahedron}.  We
have to chose a \emph{base edge} at which we start and finish the
surface composition. In our tetrahedron example this was the edge
$(14)$. When we change the base edge, holding its two end points
fixed, then the $2$-holonomy $h^\prime$ of such a closed surface
(\emph{Wilson surface}) is conjugated,
\begin{equation}
\label{eq_surfacecov1}
  h^\prime\mapsto hh^\prime h^{-1},
\end{equation}
by the label $h\in H$ associated with the surface enclosed between the
old and the new base edge. If we change the reference point of a
closed surface (the starting point of its base edge) by whiskering
with $g\in G$, then the $2$-holonomy is acted upon by the
corresponding parallel transport,
\begin{equation}
\label{eq_surfacecov2}
 h^\prime\mapsto g\rhd h^\prime.
\end{equation}

We have seen that the $2$-holonomies, \ie\ the labels $h^\prime$
associated with closed surfaces, are contained in $\ker t$. The
functions $s\colon \ker t\to \R$ that are independent of the base edge
and of the reference points, \ie\ that satisfy for all $g\in G$, $h\in
H$ and $h^\prime\in \ker t$,
\begin{eqnarray}
  s(hh^\prime h^{-1})&=&s(h^\prime),\notag\\
  s(g\rhd h^\prime)&=&s(h^\prime),
\end{eqnarray}
are called \emph{2-actions}. We have shown in~\cite{Pf03} that these
are precisely the functions of the $2$-holonomy that are invariant
under the local $2$-gauge transformations~\eqref{eq_2gauge1}
and~\eqref{eq_2gauge2}, hence the name. They form the generalization
of the Wilson action to $2$-gauge theory.

For the Euclidean and Poincar\'e $2$-groups, the $2$-actions are the
maps $s\colon H\to\R$ that are constant on the orbits of $G$ on
$H=\R^n$, \ie\ they are functions of the invariant Euclidean or
Minkowski norm, $s(v)=f(\eta(v,v))$ where $f\colon\R\to\R$ is any
function, $v\in\R^n$, and $\eta$ denotes the Euclidean or Minkowski
scalar product.  For the automorphism $2$-group, any map $s\colon
Z(H)\to\R$ gives rise to an acceptable $2$-action.

Even though there exists no canonical surface ordering, we have shown,
using ideas from the theory of $2$-categories, that the interplay of
curves and surfaces not only circumvents the no-go theorems, but also
provides us with an unambiguous and gauge covariant composition of
labeled surfaces.

Whereas in conventional gauge theory, the gauge invariant
expressions are associated with closed loops, we have seen that in
$2$-gauge theory, we can form $2$-gauge invariant expressions for
closed surfaces. Is there a also a $2$-gauge invariant expression
associated with loops, \ie\ a direct generalization of the Wilson loop
to $2$-gauge theory?

Recall that in a $1$-gauge theory, we would just calculate the (real
part of a) character,
\begin{equation}
\label{eq_1char}
  \chi(\alignidx{g_2^{-1}g_1}),
\end{equation}
of the holonomy $g_2^{-1}g_1\in G$ in order to obtain a locally
$1$-gauge invariant expression. If we take into account a possibly
non-trivial transport of curves along surfaces, then we cannot
directly compare the two curve labels $g_1$ and $g_2$, but rather have
to surface transport one curve onto the other,
\begin{equation}
\label{eq_surfacetrsp}
\begin{aligned}
\xymatrix{
  \ar@/^2ex/[rr]^{g_1}="g1"\ar@/_2ex/[rr]_{g_2}="g2"&&
  \ar@{=>}^h "g1"+<0ex,-2ex>;"g2"+<0ex,2ex>
} \qquad\longrightarrow\qquad \xymatrix{
  \ar@<.5ex>@/_2ex/[rr]^{t(h)g_1}\ar@<-.5ex>@/_2ex/[rr]_{g_2}&&
}
\end{aligned}
\end{equation}
Rather than the usual holonomy $\alignidx{g_2^{-1}g_1}$, we should
therefore consider the expression,
\begin{equation}
\label{eq_curlyf}
  \ff = \alignidx{g_2^{-1}t(h)g_1},
\end{equation}
which takes the surface transport into account. Due to
condition~\eqref{eq_constraint}, however, this expression always gives
the group unit of $G$, $\ff=1$.

There is therefore no loop based gauge invariant expression in
$2$-gauge theory which would generalize the Wilson loop of $1$-gauge
theory.

% ==============================================================================
%
\section{The differential formulation}
%
% ==============================================================================
\label{sect_differential}

In this section, we derive the differential formulation of $2$-gauge
theory which corresponds to the integral picture of the previous
section. We therefore study the integral formulation of $2$-gauge
theory on squares, cubes and hypercubes, assume that the labels depend
smoothly on the positions of the curves and surfaces and consider the
limit in which these shrink to infinitesimal size.

The theory we derive uses the same connection $1$- and $2$-forms as
Baez~\cite{Ba02} which have also been found independently by
Hofman~\cite{Ho02}, but with a flatness condition at level $1$ which
has not yet appeared in the literature. As a bonus, we can also derive
the local gauge transformations.

%------------------------------------------------------------------------------
\subsection{Lie $2$-algebras}
%------------------------------------------------------------------------------
\label{sect_2algebra}

Just as the differential picture of conventional gauge theory involves
the Lie algebra $\g$ of the gauge group $G$, we need here the
appropriate generalized notion of a `Lie algebra' associated with the
gauge $2$-group.

A \emph{Lie $2$-algebra} consists of two Lie algebras $\g$ and $\h$
with two maps\footnote{We describe here a differential crossed module,
a structure from which we can construct a strict Lie
$2$-algebra~\cite{Ba02,BaCr03}.}. The first map, $\tau\colon\h\to\g$,
is a homomorphism of Lie algebras, \ie\ a linear map that satisfies,
\begin{equation}
  \tau([Y_1,Y_2]) = [\tau(Y_1),\tau(Y_2)],
\end{equation}
for all $Y_1,Y_2\in\h$. The second map is an action of $\g$ on $\h$ by
derivations, \ie\ an operation $X\rhd Y$ for $X\in\g$, $Y\in\h$,
taking values in $\h$, such that it is an action, \ie\
\begin{equation}
  [X_1,X_2]\rhd Y = X_1\rhd(X_2\rhd Y) - X_2\rhd(X_1\rhd Y),
\end{equation}
for all $X_1,X_2\in\g$ and $Y\in\h$, and such that for any $X\in\g$,
the map $Y\mapsto X\rhd Y$ is a derivation on $\h$, \ie\
\begin{equation}
  X\rhd[Y_1,Y_2] = [X\rhd Y_1,Y_2] + [Y_1,X\rhd Y_2],
\end{equation}
for all $Y_1,Y_2\in\h$. These maps are required to satisfy the
following two compatibility conditions,
\begin{eqnarray}
  \tau(X\rhd Y) &=& [X,\tau(Y)],\label{infinitdt} \\
  \tau(Y)\rhd Y^\prime &=& [Y,Y^\prime],\label{infinitda}
\end{eqnarray}
for all $X\in\g$, $Y,Y^\prime\in\h$.

Given some Lie $2$-group in terms of the Lie groups $G$, $H$ and the
maps $t$ and $\rhd$ (Section~\ref{sect_2group}), one can construct its
Lie $2$-algebra as follows~\cite{BaCr03}. The Lie algebras $\g$ and
$\h$ are the Lie algebras of the Lie groups $G$ and $H$. The map
$\tau\colon\h\to\g$ is the derivative $\tau=dt$ of the map $t\colon
H\to G$. Finally, let the map,
\begin{equation}
  \alpha\colon G\to\Aut H,\qquad
  \alpha(g)[h] := g\rhd h,
\end{equation}
associate an automorphism $\alpha(g)$ of $H$ with each element $g\in
G$. Then the derivative of $\alpha$,
\begin{equation}
  d\alpha\colon\g\to\Der\h,\qquad X\mapsto d\alpha(X),
\end{equation}
associates with each element $X\in\g$ a derivation $d\alpha(X)$ of
$\h$. The operation $\rhd$ in the definition of the Lie $2$-algebra is
chosen to be $X\rhd Y:=d\alpha(X)[Y]$.

Consider first the Lie $2$-algebra of the Euclidean and Poincar\'e
$2$-groups. In this case, $\g=\so(n)$ or $\g=\so(n-1,1)$, and
$\h=\R^n$. The action of $\g$ on $\h$ is in the defining
representation of $\g$, and the map $\tau\colon\h\to\g$ is the null
map.

For the automorphism $2$-group of $H=\SU(2)$, we have $G=\SU(2)/\Z_2$.
In this case, both Lie algebras agree, $\g=\h$, and we have $t(Y)=Y$
for all $Y\in\h$. Finally, the action of $\g$ on $\h=\g$ is the
adjoint action, $X\rhd Y=[X,Y]$.

Let us conclude this section with a remark on the category theory
underlying the construction of Lie $2$-algebras. When we construct a
Lie 2-algebra from the differential crossed module, there is the
condition (analogous to (\ref{eq_constraint})),
\begin{equation}
\label{eq_constraint1}
  \tau (Y)= X_2-X_1, 
\end{equation}
for each $2$-cell,
\begin{equation}
\begin{aligned}
\xymatrix{
  \ar@/^2ex/[rr]^{X_1}="g1"\ar@/_2ex/[rr]_{X_2}="g2"&&
  \ar@{=>}^Y "g1"+<0ex,-2ex>;"g2"+<0ex,2ex>
}
\end{aligned}
\end{equation}
$X_i\in\g$, $Y\in\h$, of the $2$-category which is defined by the Lie
$2$-algebra. In the gauge theory language, this would correspond to an
infinitesimally small surface. The condition~\eqref{eq_constraint1} is
in fact already present in the $2$-vector spaces of~\cite{BaCr03}.

%------------------------------------------------------------------------------
\subsection{Notation}
%------------------------------------------------------------------------------
\label{sect_diff2gauge}

For the discussion of the differential picture of higher gauge theory,
we restrict ourselves to trivial bundles and present the theory in the
language of the $\g$- and $\h$-valued connection $1$- and
$2$-forms. As we will see in the following section, the basic fields
of the differential picture are a $\g$-valued connection $1$-form $A$
and an $\h$-valued connection $2$-form $B$.

We denote by $d_A$ the exterior covariant derivative for the
connection $A$ which acts on $\g$-valued $p$-forms $\phi$ by,
\begin{equation}
  d_A(\phi) = d\phi + [A,\phi].
\end{equation}
Here the bracket of a $1$-form $A$ with a $p$-form $\phi$, both taking
values in $\g$, is defined by,
\begin{equation}
  [A,\phi] := A^a\wedge\phi^b\,[T_a,T_b],
\end{equation}
where we have chosen a basis $(T_a)$ of $\g$ and written $A=A^aT_a$,
$\phi=\phi^bT_b$ with coefficient forms $A^a$ and $\phi^b$. Summation
over repeated indices is understood. Similarly, we define the action
of $d_A$ on $\h$-valued $p$-forms $\psi$, using the action of $\g$ on
$\h$ via the operation $\rhd$,
\begin{equation}
  d_A(\psi) = d\psi + A\rhd \psi,
\end{equation}
where the $\rhd$ of a $\g$-valued $1$-form with an $\h$-valued
$p$-form is defined by,
\begin{equation}
  A\rhd\psi := A^a\wedge\psi^b\,(T_a\rhd T^\prime_b),
\end{equation}
where $(T_a)$ denotes a basis of $\g$ and $(T^\prime_b)$ a basis of
$\h$. We calculate for $\g$-valued $p$-forms $\phi$,
\begin{equation}
  (d_A\circ d_A)(\phi) = [F,\phi],
\end{equation}
and for $\h$-valued $p$-forms $\psi$,
\begin{equation}
  (d_A\circ d_A)(\psi) = F\rhd\psi.
\end{equation}

%------------------------------------------------------------------------------
\subsection{Configuration variables and their curvature}
%------------------------------------------------------------------------------
\label{sect_variables}

\begin{figure}[t]
\begin{center}
\input{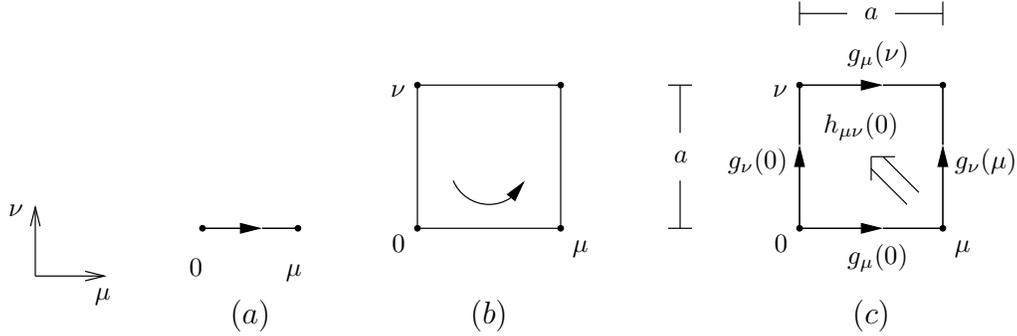}
\end{center}
\caption{
\label{fig_lattice}
Our conventions for the integral formulation of $2$-gauge theory on
the squares, in particular (c) the generalized holonomy~\eqref{eq_holonomy}.
}
\end{figure}

We shall first identify the basic fields and their associated
curvature. We then compute their respective transformations under the
local $2$-gauge transformations.

We assume that all labels depend smoothly on the curves and
surfaces. The key idea for the derivation of the differential picture
is to write down the integral formulation on squares and cubes and
then to study the limit in which these shrink to infinitesimal
size. We write the labels $g\in G$, $h\in H$ as exponentiated curve
and surface integrals over approximately constant differential forms,
\begin{eqnarray}
\label{infinit1}
\begin{array}{rcl}
g_{\mu}(0)    &=& e^{\int_{\gamma} A} \sim e^{aA_\mu},\\
h_{\mu\nu}(0) &=& e^{\int_S B}        \sim e^{a^2B_{\mu\nu}}.
\end{array}
\end{eqnarray}
Here $\gamma$ denotes a curve of length $a$ from $x=0$ to $x=\mu$ and
$S$ a square of area $a^2$ in the $(\mu\nu)$-plane. We abbreviate the
coordinates by $x=\mu:=ae_\mu$ where $e_\mu$ is a vector of unit
length. All $A_\mu$, \etc\ without argument are at $x=0$.

The basic fields in the differential picture are the $\g$-valued
connection $1$-form $A=A_\mu dx^\mu$ and the $\h$-valued connection
$2$-form $B=\frac{1}{2}B_{\mu\nu}dx^\mu\wedge dx^\nu$. Note that
$h_{\mu\nu}=h_{\nu\mu}^{-1}$.

We will make use of the usual Taylor expansion,
\begin{eqnarray}
\label{infinit2}
g_{\mu}(\alpha)    &\sim& e^{aA_{\mu}+ a^2 \partial_{\alpha} A_{\mu}},\\
h_{\mu\nu}(\alpha) &\sim& e^{a^2B_{\mu\nu}+ a^3 \partial_{\alpha} B_{\mu\nu}}.
\end{eqnarray}
When we have a product of Lie group elements, the Baker--Hausdorff
formula allows us to get the corresponding operation at the Lie
algebra level,
\begin{equation}\label{baker}
  e^xe^y= e^{x+y+\demi[x,y]+\cdots}.
\end{equation}
The action $d\alpha$ of $\g$ on $\h$ is the infinitesimal version of
the action of $G$ over $H$,
\begin{equation}
\label{infinit3}
  g_{\beta}(0)\act h_{\mu\nu}(\beta) \sim e^{a^2B_{\mu\nu}+a^3
  \partial_{\beta}B_{\mu\nu}+a^3 d\alpha (A_{\beta}) (B_{\mu\nu})}.
\end{equation}
The map $\tau\colon\h\rightarrow\g$ is the infinitesimal version of
the map $t\colon H\rightarrow G$,
\begin{equation}
  t(h_{\mu\nu})\sim e^{a^2\tau(B_{\mu\nu})}.
\end{equation}
As mentioned earlier, they satisfy the compatibility
conditions~(\ref{infinitda}) and~(\ref{infinitdt}). The approximations
(\ref{infinit1}--\ref{infinit3}) together with (\ref{infinitdt},
\ref{infinitda}) are all we need in order to to derive the
differential picture.

So far we have identified as the basic fields the generalized
connection $(A,B)$ in agreement with~\cite{Ba02}. Let us now calculate
a curvature $2$-form, using the holonomy around an infinitesimal
square, and a curvature $3$-form, using the $2$-holonomy around an
infinitesimal cube.

% These two should be scaled down to 75% already in xfig before exporting!
\begin{figure}[t]
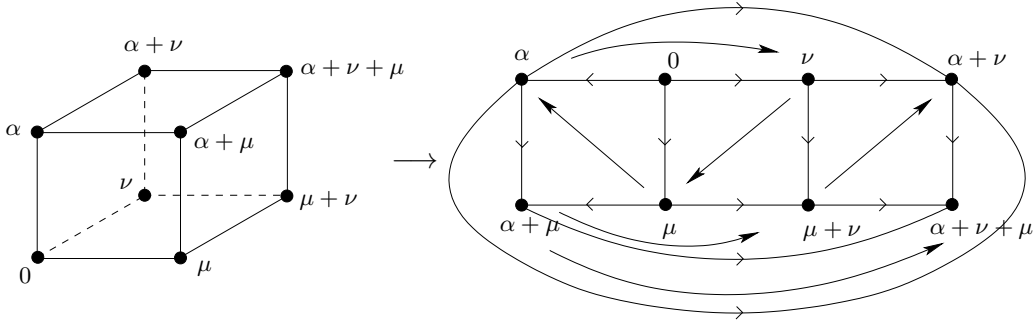

\begin{center}
\begin{minipage}[c]{5cm}
\input{cube.pstex_t}
\end{minipage}
$\longrightarrow$
\begin{minipage}[c]{7.8cm}
\input{flatcube.pstex_t}
\end{minipage}
\caption{
\label{flatcube}
A flattened cube in order to read off the $2$-holonomy~\eqref{2holonomy}.
}
\end{center}
\end{figure}

In $1$-gauge theory, the curvature $2$-form is given by an
infinitesimal Wilson loop. In the context of $2$-gauge theory, it
depends also on the $B$-field because of~\eqref{eq_curlyf}. The
expression $\ff$ of~\eqref{eq_curlyf} reads for the square of
Figure~\ref{fig_lattice}(c),
\begin{equation}
\label{eq_holonomy}
  e^{a^2\tilde F_{\mu\nu}} \sim \ff _{\mu\nu} = g_{\mu}^{-1}(\nu)g_{\nu}(0)\mone t(h_{\mu\nu}(0))g_{\mu}(0)g_{\nu}(\mu).
\end{equation}
Using the approximations~(\ref{infinit1}--\ref{baker}) and dropping
all terms of order $a^3$ in the exponent, we obtain the curvature
$2$-form $\tilde F_{\mu\nu}=\frac{1}{2}\tilde F_{\mu\nu}dx^\mu\wedge
dx^\nu$ as follows,
\begin{eqnarray}
  \tilde F_{\mu\nu} &=& \partial_{\mu}A_{\nu}- \partial_{\nu}A_{\mu}+
                        [A_{\mu},A_{\nu}]+  \tau(B_{\mu\nu})\nn\\
                    &=& F_{\mu\nu} + \tau(B_{\mu\nu}),
\end{eqnarray}
where $F$ denotes the conventional curvature of $A$. This agrees with
Baez' expression~\cite{Ba02} (modulo a sign which is matter of
convention),
\begin{equation}
  \tilde F = dA + \frac{1}{2}[A, A] + \tau(B) = F + \tau(B).
\end{equation}
We have therefore found a geometrical interpretation for the
generalized curvature $\tilde F$ from the surface transport of the
curve~\eqref{eq_surfacetrsp}.

It is important here to remember that in the definition of a strict
Lie $2$-group, we have $\ff=1\in H$ in~\eqref{eq_curlyf} and therefore
at the differential level,
\begin{equation}
\label{eq_integrability}
  \tau(B)=-F.
\end{equation}
This condition is the alter ego of the equations (\ref{eq_constraint})
and~\eqref{eq_constraint1} and has some drastic consequences: it means
that the curvature $2$-form $\tilde F$ is always zero.

Let us now compute the curvature $3$-form
$G=\frac{1}{6}G_{\alpha\mu\nu}dx^\alpha\wedge dx^\mu\wedge dx^\nu$
associated with $A$ and $B$. It is the differential counterpart of the
$2$-holonomy around a cube (Figure~\ref{flatcube}).

To calculate it, we use the same technique as for the tetrahedron in
(\ref{eq_tetrahedron}, \ref{eq_tetcomposite}). We obtain,
\begin{eqnarray}
\label{2holonomy}
e^{a^3 G_{\alpha\mu\nu}} &\sim& \gg_{\alpha\mu\nu}\nn\\
  &=& [g_{\alpha}(0)\act
       h_{\mu\nu}(\alpha)]h_{\mu\alpha}(0)[g_{\mu}(0) \act
       h_{\nu\alpha}(\mu)]h_{\nu\mu}(0)[g_{\nu}(0)\act h_{\alpha\mu}(\nu)]
       h_{\alpha\nu}(0).
\end{eqnarray}
Let us use once again the the approximations
(\ref{infinit1}--\ref{baker}) and drop all terms of order $a^4$ in the
exponent, so that we get,
\begin{equation}
  G_{\alpha\mu\nu}
    = \partial_{\alpha}B_{\mu\nu}+ d\alpha (A_{\alpha})(B_{\mu\nu})+\partial_{\nu}B_{\alpha\mu}+
        d\alpha(A_{\nu})(B_{\alpha\mu})+\partial_{\mu}B_{\nu\alpha}+
        d\alpha(A_{\mu})(B_{\nu\alpha}),
\end{equation}
and using the simplified notation,
\begin{equation} 
\label{eq_curvature3}
  G= dB + A\rhd B = d_A(B).
\end{equation}
This coincides with Baez' definition of the curvature
$3$-form~\cite{Ba02}.

%------------------------------------------------------------------------------
\subsection{`Differential' gauge transformations}
%------------------------------------------------------------------------------
\label{localgauge}

In order to derive the differential form of the local $2$-gauge
transformations~\eqref{eq_2gauge1} and~\eqref{eq_2gauge2}, we draw the
analogous diagrams for a square and a cube, respectively, \cf\
Figure~\ref{fig_lattice} and~\ref{flatcube}. Here the old
configuration corresponds to the bottom of the diagram, the new one to
the top. We parameterize `differential' gauge transformations by the
height $\epsilon$ of these diagrams, \ie\ the $2$-generating function,
\begin{eqnarray}
  \eta_{\alpha}(0)   &\sim& e^{\epsilon X}, \\
\label{eq_gaugedefy}
  \eta_{\mu\alpha}(0)&\sim& e^{\epsilon aY_{\mu}},
\end{eqnarray}
is parameterized by a $\g$-valued function $X$ and by an $\h$-valued
$1$-form $Y=Y_\mu dx^\mu$. Similarly to Section~\ref{sect_variables},
we use the Taylor expansion,
\begin{eqnarray}
  \eta_\alpha(\mu)      &\sim& e^{\epsilon(X + a\del_\mu X)},\\
  \eta_{\mu\alpha}(\nu) &\sim& e^{\epsilon(aY_\mu + a^2\del_\nu Y_\mu)},
\end{eqnarray}
the convention $\eta_{\mu\alpha}=\eta_{\alpha\mu}^{-1}$, the
derivative $\tau=dt$,
\begin{equation}
  t(\eta_{\mu\alpha}(0)) \sim e^{a\epsilon\tau(Y_\mu)},
\end{equation}
and the group actions,
\begin{eqnarray}
  g_\nu(0)\rhd\eta_{\mu\alpha}(\nu) &\sim& e^{\epsilon(aY_\mu+a^2\del_\nu Y_\mu+a^2d\alpha(A_\nu)(Y_\mu))},\\
  \eta_\alpha(0)\rhd h_{\mu\nu}(0) &\sim& e^{a^2(B_{\mu\nu}+\epsilon d\alpha(X)(B_\mu\nu))}.
\end{eqnarray}

The gauge transformation for the connection $1$-form $A$ is read off
from the square of Figure~\ref{fig_lattice} and the formula
(\ref{eq_1gauge}),
\begin{equation} 
\label{Altransf}
  e^{aA_\mu + a\epsilon\delta A_\mu} \sim {\tilde g}_{\mu}(\alpha) =
  \alignidx{\eta_{\alpha}^{-1}(0)t(\eta_{\mu\alpha}(0))g_{\mu}(0)\eta_{\alpha}(\mu)}.
\end{equation}
Using the approximations (\ref{infinit1}--\ref{baker}) and dropping
terms of order $\epsilon^2$ and $a^2$ in the exponent, we get the
gauge transformation,
\begin{equation}
\label{Atransf}
  A_{\mu}\mapsto A_{\mu} +\epsilon\delta A_\mu,\qquad
  \delta A_\mu = \partial_{\mu}X + [A_{\mu}, X] + \tau(Y_{\mu}),
\end{equation}
that is
\begin{equation} 
\label{Altransf1}
  \delta A=d_A(X) + \tau(Y).
\end{equation}

The 2-gauge transformations of the $B$-field can be deduced from the
flattened cube (Figure~\ref{flatcube}) whose height in
$\alpha$-direction is $\epsilon$ for the gauge transformation,
\begin{eqnarray}
\label{Bltransf}
  e^{a^2B_{\mu\nu}+a^2\epsilon\delta B_{\mu\nu}} &\sim& \tilde h_{\mu\nu}(\beta)\\
  &=& \eta_{\alpha}(0)\mone \act \left\{
      \eta_{\nu\alpha}(0) [g_{\nu}(0)\act \eta_{\mu\alpha}(\nu)]
      h_{\mu\nu}(0) [g_{\mu}(0)\act
      \eta_{\nu\alpha}^{-1}(\mu)]\eta_{\mu\alpha}^{-1}(0)\right\}.\nn
\end{eqnarray}
The infinitesimal transformation is calculated as usual with the help
of (\ref{infinit1}--\ref{baker}), dropping terms of order $\epsilon^2$
and $a^3$ in the exponent, and we get,
\begin{eqnarray}
\label{Btransf}
  &B_{\mu\nu}\mapsto B_{\mu\nu} + \epsilon\delta B_{\mu\nu},&\nn\\
  &\delta B_{\mu\nu} = 
  -\del_\mu Y_\nu + \del_\nu Y_\mu - d\alpha(A_\mu)(Y_\nu) + d\alpha(A_\nu)(Y_\mu) - d\alpha(X)(B_{\mu\nu}).&
\end{eqnarray}
So by using the shorthand notation, we have
\begin{equation} 
\label{Bltransf1}
  \delta B = - dY - A\rhd Y - X\rhd B = - d_A(Y) - X\rhd B.
\end{equation}

From the gauge transformations~\eqref{Atransf} and~\eqref{Btransf} for
$A$ and $B$, we can deduce the transformation of the curvature
$2$-form,
\begin{equation}
  \tilde F\mapsto \tilde F + \epsilon\delta\tilde F,\qquad
  \delta\tilde F = [\tilde F,X],
\end{equation}
up to terms of order $\epsilon^2$. This means that $\tilde F$
transforms covariantly even without the assumption $\tilde F=0$, and
on the other hand that the transformation preserves the condition
$\tilde F=0$.

For the curvature $3$-form $G$, we obtain
after a rather lengthy calculation,
\begin{equation}
  G\mapsto G+\epsilon\delta G,\qquad
  \delta G = -\tilde F\rhd Y - X\rhd G,
\end{equation}
up to terms of order $\epsilon^2$. This transformation shows that $G$
transforms covariantly if and only if $\tilde F=0$. However, if we are
considering the case of a strict Lie $2$-algebra, then this flatness
condition is naturally present, and $G$ transforms covariantly,
moreover it sees only the level $1$ of the generating function,
\begin{equation}
  \delta G = - X\rhd G.
\end{equation}

%------------------------------------------------------------------------------
\subsection{`Large' gauge transformations}
%------------------------------------------------------------------------------

In the previous section, we have derived the `differential' gauge
transformations in the differential picture. What can we conclude from
their existence?

Recall first the role of `large' and `differential' gauge
transformations in conventional non-Abelian gauge theory with gauge
group $G$. A `large' gauge transformation is a bundle automorphism of
the principal bundle $P\to M$. In a local trivialization on
$U\subseteq M$, it is given by a $G$-valued generating function
$g\colon U\to G$.  The connection $1$-form and the curvature $2$-form
transform as,
\begin{eqnarray}
\label{eq_gaugelargea}
  A &\mapsto& g^{-1}Ag + g^{-1}dg,\\
  F &\mapsto& g^{-1}Fg.
\end{eqnarray}
It is often convenient to consider only the tangents to the above
transformations which means to parameterize $g$ in terms of the Lie
algebra,
\begin{equation}
\label{eq_gaugesmall}
  g = e^{\epsilon X},
\end{equation}
where $X\colon U\to\g$ is a Lie algebra valued function. If $G$ is
compact and connected, then the exponential map is surjective, see,
for example~\cite{Co84}, and any generating function is of this
form. If $G$ is non-compact or not connected, this is in general no
longer true. Usually, one considers the
parameterization~\eqref{eq_gaugesmall} only for small $\epsilon$ and
finds,
\begin{eqnarray}
  A &\mapsto& A + \epsilon(dX + [A,X]) = A + \epsilon d_A(X),\\
  F &\mapsto& F + \epsilon [F,X],
\end{eqnarray}
dropping terms of order $\epsilon^2$. Since we know that we can always
integrate these `differential' gauge transformations, we can recover
the `large' transformations as long as the exponential map is
surjective.

\begin{figure}[t]
\begin{center}
\input{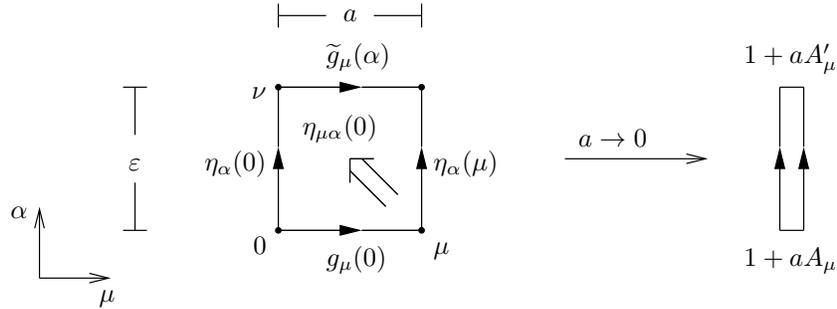}
\end{center}
\caption{
\label{fig_largegauge}
The local gauge transformation of the edge labels in the integral
picture, \cf~\eqref{eq_2gaugelarge}. The bottom layer is the old
configuration, the top layer the new one. In order to pass to the
differential picture, we shrink the rectangle to infinitesimal width,
$a\rightarrow 0$, but keep its height $\epsilon$ fixed.}
\end{figure}

Let us now try to derive the `large' counterparts of the
`differential' $2$-gauge transformations of
Section~\ref{localgauge}. Therefore, we again consider the integral
formulation on squares and cubes, but this time we keep the height
$\epsilon$ fixed and consider only the limit $a\to 0$, see
Figure~\ref{fig_largegauge}.

\begin{figure}[t]
\begin{center}
\input{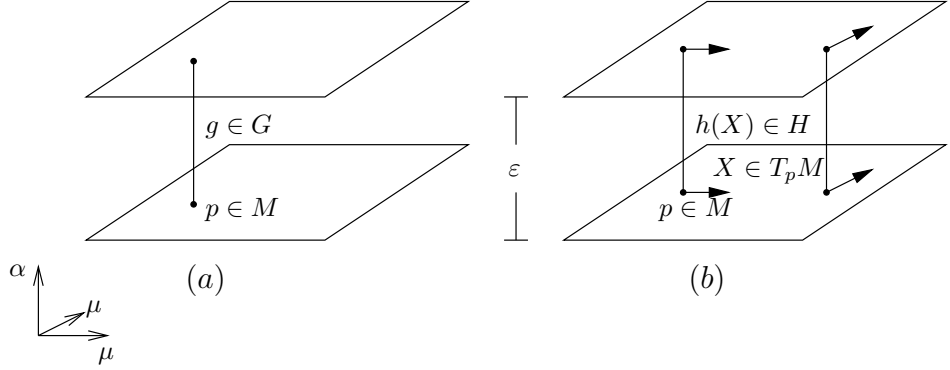}
\end{center}
\caption{
\label{fig_wishful}
(a) After shrinking the square of Figure~\ref{fig_largegauge} to
infinitesimal width $a\rightarrow 0$, its vertical edges carry a group
label $g(p)=\eta_\alpha(p)\in G$ at each point $p\in M$. (b) The
vertical surface label $\eta_{\mu\alpha}(p)$ would ideally yield an
$H$-valued $1$-form $h_\mu(p)dx^\mu=\eta_{\mu\alpha}(p)dx^\mu$ which
associates with each vector $X\in T_pM$ an element $h(X)\in H$. The
linear structure of $T_pM$ here imposes a serious constraint as we
explain in the text.
}
\end{figure}

First we restrict ourselves to the case in which the curve labels of
the gauge generating transformation are trivial, \ie\ the surface in
Figure~\ref{fig_largegauge} has the trivial label
$\eta_{\mu\alpha}(0)=1\in H$. Rather than~\eqref{Altransf}, we now
write $e^{aA_\mu}\sim g_\mu(0)$ and $e^{aA^\prime_\mu}\sim \tilde
g_\mu(\alpha)$ and obtain,
\begin{equation}
\label{eq_2gaugelarge}
  (1+aA_\mu^\prime)\sim e^{aA_\mu^\prime} \sim \eta_\alpha(0)^{-1}
  e^{aA_\mu}\eta_\alpha(\mu) \sim
  \eta_\alpha(0)^{-1}(1+aA_\mu)(\eta_\alpha(0) + a\del_\mu\eta_\alpha(0)).
\end{equation}
Dropping terms of order $a^2$, this gives the familiar transformation
rule~\eqref{eq_gaugelargea} for the $G$-valued function
$g(p)=\eta_\alpha(p)$, $p\in U$. The index $\alpha$ was just used in
order to denote the vertical `gauge' direction in our figure.

All the data required in order to describe the gauge transformation
are associated with vertical curves or surfaces which link the bottom
with the top layer of Figure~\ref{fig_largegauge}. If the surface
label is trivial, \ie\ if the gauge generating transformation assigns
the group unit $1\in H$ to each curve, we have to deal only with
vertical lines labeled by elements,
\begin{equation}
\label{eq_2gaugelargeg}
  \eta_\alpha(p)=g(p)\in G,
\end{equation}
at each point $p\in U$, Figure~\ref{fig_wishful}(a). Indeed, the gauge
generating function,
\begin{equation}
\label{eq_2gaugelargeg2}
  g\colon U\to G,
\end{equation}
can be visualized by a bunch of such vertical lines between the old
configuration (bottom) and the new one (top). It poses no problem that
the labels are in the group $G$. The transformation is just a change
of coordinates.

Let us now consider the case in which the curve labels of the
$2$-generating function are non-trivial, \ie\ the square in
Figure~\ref{fig_largegauge} has a non-trivial label
$\eta_{\mu\alpha}(0)\in H$. The index $\alpha$ just indicates that the
surface is vertical, but the index $\mu$ has indeed a geometrical
meaning. In order to derive the `differential' gauge transformations
we have expanded~\eqref{eq_gaugedefy} in terms of both $a$ and
$\epsilon$ in order to obtain a $1$-form $Y_\mu dx^\mu$ as the
differential expression.

If we expand only in terms of $a$, but keep $\epsilon$ fixed, we run
into the following geometrical obstruction as depicted in
Figure~\ref{fig_wishful}(b). By expansion in terms of $a$, we wish to
obtain a $1$-form, say $h_\mu dx^\mu$, from the vertical surface. For
any tangent vector $X\in T_pM$ at some point $p\in U$, we should
therefore be able to evaluate $h(X)=h_\mu X^\mu$. On the other hand,
from the surface label, $\eta_{\mu\alpha}(0)\in H$, there remains for
each choice of $\mu$ a group element in $H$. We are tempted to write,
\begin{equation}
  \eta_{\mu\alpha}(p) = h_\mu(p)\in H.
\end{equation}
Since $T_pM$ is a linear space, however, this is possible only if
$h_\mu$ gives rise to a \emph{linear} map,
\begin{equation}
  h\colon TM\to H.
\end{equation}
This condition is stronger than that in~\eqref{eq_2gaugelargeg}
and~\eqref{eq_2gaugelargeg2}.

The construction of the `large' gauge transformations in the
differential picture is therefore possible only if $H=\R^n$ for some
$n$. In this case, the action $\rhd$ of $G$ on $H$ is actually a
representation of the Lie group $G$ and induces a representation
$\rhd$ of the Lie algebra $\g$ on $H$.

We parameterize the `large' gauge transformations by $h_\mu(p)\in H$,
$p\in M$, \ie\
\begin{equation}
\label{eq_largegaugeh}
  \eta_{\mu\alpha}(p) = ah_\mu(p),
\end{equation}
identifying $H$ with its Lie algebra and indicating that $h_\mu$ is
already a quantity of order $a$. Rather than~\eqref{Altransf}, we
obtain for the transformation of the connection $1$-form,
\begin{equation}
\label{eq_2gaugelarge1}
  A \mapsto A^\prime = g^{-1}A g + g^{-1}dg + \tau(h).
\end{equation}
In order to derive the `large' gauge transformations for the
connection $2$-form $B_{\mu\nu}$, we replace~\eqref{Bltransf} by,
\begin{equation}
  e^{a^2B_{\mu\nu}^\prime}\sim 
  \eta_\alpha^{-1}(0)\rhd\biggl[
    \eta_{\nu\alpha}(0)(e^{aA_\nu}\rhd\eta_{\mu\alpha}(\nu))e^{a^2B_{\mu\nu}}
    (e^{aA_\mu}\rhd\eta_{\nu\alpha}^{-1}(\mu))\eta_{\mu\alpha}^{-1}(0)\biggr].
\end{equation}
Expanding everything up to order $a^2$, using $\eta_\alpha(0)=g(0)$,
\eqref{eq_largegaugeh}, and $h_\mu(\nu)=h_\mu(0)+a\del_\nu h_\mu(0)$,
we obtain the transformation,
\begin{equation}
\label{eq_2gaugelarge2}
  B\mapsto B^\prime = g^{-1}\rhd B - d_A(h).
\end{equation}
From the `large' gauge transformations~\eqref{eq_2gaugelarge1}
and~\eqref{eq_2gaugelarge2} one can recover the `differential'
transformations~\eqref{Altransf1} and~\eqref{Bltransf1} using the
parameterizations~\eqref{eq_gaugesmall} and~\eqref{eq_largegaugeh}.

We conclude that we have a full extended local gauge symmetry only if
$H\cong\R^n$. In general, it is not possible to integrate the
`differential' gauge transformations and to obtain proper (`large')
transformations. The `large' transformations are given
by~\eqref{eq_2gaugelarge1}, \eqref{eq_2gaugelarge2} and,
\begin{equation}
  G\mapsto G^\prime =g^{-1}\rhd G.
\end{equation}

%------------------------------------------------------------------------------
\subsection{Pure $2$-gauge and $2$-flatness}
%------------------------------------------------------------------------------

For the case $H\cong\R^n$, we can now express the condition of being
pure $2$-gauge (Section~\ref{sect_puregauge}) in the differential
language. A generalized connection $(A,B)$ is pure gauge if there
exists (locally) a $G$-valued function $g\colon U\to G$ and an
$H$-valued $1$-form $h\colon T^*M|_U\to H$ such that,
\begin{eqnarray}
  A &=& g^{-1}dg + \tau(h),\\
  B &=& -dh.
\end{eqnarray}
It is straightforward to show that these configurations are also
$2$-flat.

%------------------------------------------------------------------------------
\subsection{Flatness at level $1$}
%------------------------------------------------------------------------------

The level-$1$ flatness condition $\tilde F=0$,
\cf~\eqref{eq_integrability}, has the following effect on the
curvature $3$-form. The definition $G=d_A(B)$ implies
$\tau(G)=d_A(\tau(B))=-d_A(F)=0$ by the Bianchi identity for the
conventional curvature $F$ of $A$. This implies that $G$ takes values
in $\ker\tau\unlhd\h$ which is an Abelian ideal of $\h$. This is the
differential counterpart of the result~\cite{Pf03} that the
$2$-holonomy of an $S^2$ surface in the integral picture takes values
in the Abelian normal subgroup $\ker t\unlhd H$
(Section~\ref{sect_2gaugeinv}).

We can always decompose the $\h$-valued connection $2$-form $B=z\oplus
B^\prime$ where $z\in\ker\tau$ and $\h=\ker\tau\oplus\h^\prime$ is
split into a direct sum of vector spaces. The non-Abelian part of $B$
is therefore contained in $B^\prime\in\h^\prime$ and related to the
conventional curvature $F$ of $A$ by $F=-\tau(B^\prime)$ due
to~\eqref{eq_integrability}. The only contribution to $B$ unrelated to
the curvature of $A$ is contained in some Abelian sub-algebra of $\h$.

%------------------------------------------------------------------------------
\subsection{Higher Bianchi identities}
%------------------------------------------------------------------------------

As a generalization of the Bianchi identity $d_A(F)=0$ of conventional
$1$-gauge theory, we have in $2$-gauge theory,
\begin{equation}
  d_A(\tilde F) = d_A(F) + \tau(d_A(B)) = \tau(G) = 0,
\end{equation}
where we have used the condition~\eqref{eq_integrability} only in the
last step, and
\begin{equation}
  d_A(G) = F\rhd B = -\tau(B)\rhd B = -[B,B] = 0,
\end{equation}
where we have used~\eqref{eq_integrability} in the second step. They
can be derived from the integral picture by drawing the square and
cube from the definition of the curvature $2$- and $3$-form and by
parallel transporting the entire diagram in an independent direction.

%------------------------------------------------------------------------------
\subsection{Examples}
%------------------------------------------------------------------------------

%------------------------------------------------------------------------------
%\subsection{$2$-gauge invariant actions}
%------------------------------------------------------------------------------
\label{sect_lagrangians}

\paragraph{$BF$-theory.}

Consider first the special example in which we use the adjoint
$2$-group of some Lie group $G$, \ie\ $H=\g$ is the Lie algebra, $G$
acts on $H=\g$ by the adjoint action, and the map $t\colon H\to G$ is
$t(h)=1$ for all $h\in H$. The corresponding Lie $2$-algebra is given
by the Lie algebras $\g=\h$, the adjoint action of $\g$ on $\h$ and
the null map $\tau=dt=0$.

In this case, the differential forms $A$, $B$, $F$, $\tilde F$ and $G$
are all $\g$-valued. In four dimensions, one can therefore consider
the Lagrangian of $BF$-theory~\cite{Ho89},
\begin{equation}
  \mathcal{L} = \tr_\g (B\wedge F).
\end{equation}
The local $2$-gauge transformations are generated by a $\g$-valued
function $X$ and a $\g$-valued $1$-form $Y$,
\begin{eqnarray}
  \delta A &=& d_A(X),\\
  \delta B &=& d_A(Y) - [X,B],\\
  \delta F &=& -[X,F],\\
  \delta G &=& -[X,G].
\end{eqnarray}
They encompass both the ordinary local gauge symmetry (generated by
$X$) and the extended, so-called topological, local symmetry which is
a special feature of $BF$-theory (generated by $Y$). Both are unified
in the local $2$-gauge symmetry. We have therefore discovered the
actual geometrical reason for the topological symmetry of
$BF$-theory. Notice that the level-$1$ flatness
condition~\eqref{eq_integrability} reads in this case $F=0$ which is
actually one of the field equations of $BF$-theory.

Notice that in the case of $BF$-theory, the Abelian ideal is $\ker
t=\g$ which is an Abelian group using the addition of elements of
$\g$, even though $\g$ as a Lie algebra can be non-Abelian if the
gauge group $G$ is non-Abelian.

\paragraph{Yang--Mills theory.}

Let us now try to construct a higher level analogue of the Yang--Mills
action. In conventional gauge theory, the Yang--Mills Lagrangian
reads,
\begin{equation}
  \mathcal{L} = \tr_\g (F\wedge *F),
\end{equation}
where $\tr_\g$ denotes the Cartan--Killing form of $\g$. A candidate
for a Lagrangian density in higher gauge theory is therefore given by
the expression,
\begin{equation}
\label{eq_2lagrangian}
  \mathcal{L} = \tr_\h (G\wedge *G).
\end{equation}
We could have tried $\tr_\g(\tilde F\wedge*\tilde F)$ which, however,
vanishes because of~\eqref{eq_integrability} in the case of strict Lie
$2$-groups. We have seen that the curvature $3$-form $G$ is always
Abelian.

If we choose the Euclidean or Poincar\'e $2$-group, we have
$\g=\so(n)$ or $\so(n-1,1)$, $\h=\R^n$ and $\tau\colon\h\to\g$ the
null map. This implies in particular that $\tilde F=F$, and the
condition~\eqref{eq_integrability} furthermore states that the
connection $A$ is flat.  In addition, we have a connection $2$-form
$B$ taking values in $\h=\R^n$ with a curvature $3$-form
$G=d_A(B)$. Locally, the flatness of $A$ implies that it is pure
gauge, \ie\ gauge equivalent to $A=0$, so that locally $G$ is just the
exterior derivative, $G=dB$. The Yang--Mills
Lagrangian~\eqref{eq_2lagrangian} therefore agrees locally with that
of Abelian $2$-form electrodynamics.

A similar result can be shown for all $2$-groups in which $H=V$ is a
vector space on which $G$ is represented. The connection $1$-form $A$
is locally pure gauge and the Yang--Mills
Lagrangian~\eqref{eq_2lagrangian} reduces to that of Abelian $2$-form
electrodynamics.

Consider finally the automorphism $2$-group of $H=\SU(2)$, \ie\
$\g=\h=\su(2)$, $\tau$ is the identity map, and $\g$ acts on $\h=\g$
by the adjoint action. In this case, the
condition~\eqref{eq_integrability} implies that $B=-F$ is just (minus)
the ordinary curvature $2$-form of $A$. The Yang--Mills
Lagrangian~\eqref{eq_2lagrangian} therefore vanishes because of the
conventional Bianchi identity $G=d_A(B)=-d_A(F)=0$.

With the known Lie $2$-groups alone, it is therefore not possible to
find a non-trivial generalization of the Yang--Mills action. This is
in outright contrast to the integral picture for which we have
shown~\cite{Pf03} that non-trivial generalizations exist. This result
points towards a genuine discrepancy between perturbative and
non-perturbative formulations of higher gauge theory on which we
comment in the conclusion.

% ==============================================================================
%
\section{Conclusion and Outlook}
%
% ==============================================================================
\label{sect_conclusion}

In this article, we have reviewed both the integral and the
differential picture of higher gauge theory. One main result is the
appearance of the condition~\eqref{eq_integrability} at the
differential level as soon as the curve and surface labels depend
smoothly on the positions of the curves and surfaces.

Another main result is that we are able to construct `large' (as
opposed to `differential') $2$-gauge transformations in the
differential picture only in the case in which $H\cong\R^n$ as an
Abelian group. This seriously restricts the applications of the
differential formulation and prevents us from obtaining an interesting
level-$2$ generalization of Yang--Mills theory. $BF$-theory, however,
forms an interesting example of a $2$-gauge theory. The local
$2$-gauge transformations unify the two types of local symmetries of
$BF$-theory and thereby provide a structural explanation for the
existence of the topological symmetry of $BF$-theory.

We have chosen the language of $2$-groups~\cite{Ba02} in order to
study higher gauge theory. The categorical structure of $2$-groups
leads directly to the integral picture~\cite{Pf03} and as a
consequence to the differential formulation as derived in the present
article. Alternatively it would be possible to use the language of
gerbes and to start with a differential formulation of higher gauge
theory. One can then ask under which conditions it is possible to
integrate the connection $1$-form along curves and the connection
$2$-form along surfaces in a consistent way. The result of the present
article suggests the conjecture that~\eqref{eq_integrability} is the
required integrability condition.

How serious are the restrictions we have found, in particular the
Abelianness of the curvature $3$-form?

First note that all the Lie $2$-groups and Lie $2$-algebras used in
the present article are \emph{strict}. They form only the simplest
examples of these structures which can be constructed in a general
$2$-categorical framework, but there exist the more general notions of
\emph{weak} and \emph{coherent} $2$-groups and their Lie
$2$-algebras. For $2$-groups, see, for example~\cite{La02,BaLa03} and
for Lie $2$-algebras~\cite{BaCr03}. One can hope that the differential
picture becomes less restrictive once we generalize from strict Lie
$2$-groups and $2$-algebras to weak ones. Since the origin of the
level-$1$ flatness is the very basic condition~\eqref{eq_constraint},
a fully successful weakening should therefore allow for a non-Abelian
kernel of the map $t\colon H\to G$.

We also observe that the non-trivial $2$-holonomies are ruled out in
the differential picture only if one requires the connection $1$- and
$2$-forms to be both continuous and well defined everywhere in
space-time. In particular, in the integral picture with the
automorphism $2$-group of $\SU(2)$, we can have $\Z_2$-valued
$2$-holonomies associated with surfaces of topology $S^2$. If we
assume that a smooth deformation of the surface changes the
$2$-holonomy only smoothly, then the non-trivial $\Z_2$-element
indicates that the $S^2$ is not smoothly contractible. This can be
interpreted as an indication that there are singularities of
codimension $2$ in the theory which are actually \emph{predicted} by
the algebraic structure. Soliton-like solutions of some classical
field equations come to mind. In fact, the integral picture for the
inner automorphism group of $\SU(3)$ is related to the symmetries of
centre vortices in QCD as sketched in~\cite{Pf03}.

The difference of the differential and the integral picture is much
deeper, though. As an illustration, we refer to a result in the
context of the path integral quantization of conventional gauge
theory. For simplicity, assume that we work in the Euclidean setting
(\ie\ with `imaginary' time) on some Riemannian manifold $M$.

The obvious naive choice is to consider the set $\mathcal{A}$ of all
smooth connections $A$ on $M$ which form an affine space, and then to
divide out the action of the gauge transformations. This step,
however, destroys the linear structure so that the standard techniques
fail to construct a useful path integral measure on the quotient
$\mathcal{A}/\mathcal{G}$. This failure to implement the gauge
symmetry correctly can be seen as a main reason why perturbative QCD
does not predict confinement as observed in Nature.

The sophisticated approach, see, for example~\cite{AsLe95}, is to
consider the collection of all graphs embedded in $M$, to study gauge
theory in the integral picture on these graphs, \ie\ all `connections'
that are given by group labels attached to the edges of the graph, and
finally to make use of a refinement relation on the class of all
graphs which facilitates the construction of a projective continuum
limit for the set of connections. Not only does this set of
generalized connections form a compact Hausdorff space, it is also
possible to fully divide out the gauge symmetry. This set of
generalized connections modulo gauge transformations is a huge space
that includes not only smooth or continuous connections, but rather
mainly distributional ones. In order to appreciate the physical
significance of this space of generalized connections it is useful to
recall the most basic example of a field theory which admits a
rigorous Euclidean path integral quantization, the free relativistic
scalar field. Its path integral measure~\cite{GlJa87} is supported
mainly on non-continuous scalar fields. In fact, the subsets of
continuous and smooth fields form sets of measure zero!

This is a strong indication that a restriction to smooth fields does
not yield an adequate description of the corresponding quantum theory
and that we should take the integral formulation seriously. In a
proper continuum limit, constructed from a suitable refinement of the
integral formulation of higher gauge theory, the above-mentioned
codimension-$2$ singularities will not only be allowed, they may
actually be abundant in the path integral.

%------------------------------------------------------------------------------
\acknowledgments
%------------------------------------------------------------------------------

The authors would like to thank John Baez, Louis Crane, Marni
Sheppeard, Romain Attal, Laurent Freidel and Oliver Winkler for
various stimulating discussions and Amitabha Lahiri for valuable
correspondence.

\end{document}